\documentclass{moriond}

\def\be{\begin{equation}}
\def\ee{\end{equation}}
\def\bea{\begin{eqnarray}}
\def\eea{\end{eqnarray}}

\newcommand{\Photo}{}

\begin{document}
\vspace*{4cm}
\title{Measurement of stray light in the LISA instrument}

\author{ M. Nardello \footnote{marco.nardello@oca.eu; phone +33 06 52 57 52 05; https://artemis.oca.eu/fr/accueil-artemis}, A. Roubeau-Tissot, M. Lintz }

\address{ARTEMIS (Université Cote d'Azur et Observatoire de la Côte d'Azur et CNRS),\\ 96 Boulevard de l’Observatoire, 06300 Nice
}
\maketitle\abstracts{
Measurement of stray light in a complex optical system can be a complex task. We developed a method to measure coherent stray light inside an assembled device, determining all stray light sources and their relative contribution. The method is based on the insertion of a laser with a scanned frequency and the detection of all the electrical and optical signals obtained from the instrument. The spectra calculated from these signals show fringes due to interference between each stray light contribution and the nominal beam. The frequency of these interference peaks indicates the difference in path length between the stray light path and the nominal path. To have a description of the measured data we realized optical simulations, which link the measured path length to a possible path throughout the system. In the following, we will show some measurements made on a test bench realized to simulate the performance of LISA interferometers and describe how accurate simulations are in predicting and explaining the measured results.}

\section{Introduction}

The LISA (Laser Interferometer Space Antenna) mission has the objective of detecting gravitational waves with heterodyne interferometry between three spacecraft orbiting the Sun in a triangular formation of 2.5Mkm~\cite{lisa}. Since stray light will affect the measurements, the SL-OGSE instrument (Stray Light Optical Ground Support Equipment~\cite{slogse}) has been developed to identify and measure all stray light (SL) contributions present on its optical benches.

\section{Instrument description}

The instrument uses the FMCW (Frequency Modulated Continuous Wave) technique to obtain for each SL contribution its optical path length difference (OPD) and fractional amplitude, with respect to the nominal beam. The technique consists in injecting a laser ramped in frequency on a 2~nm range around 1064.5~nm, collecting the signals from all photo-receivers, and applying a Fourier transform.
Each SL contribution will appear, in the Fourier spectrum, as a peak at a position given by Eq.~\ref{eq:freq}:

\begin{equation}
f_{SL} = \frac{OPD}{c}\frac{\Delta\nu}{\Delta t}
\label{eq:freq}
\end{equation}

\begin{figure}
\centerline{\includegraphics[width=0.65\linewidth]{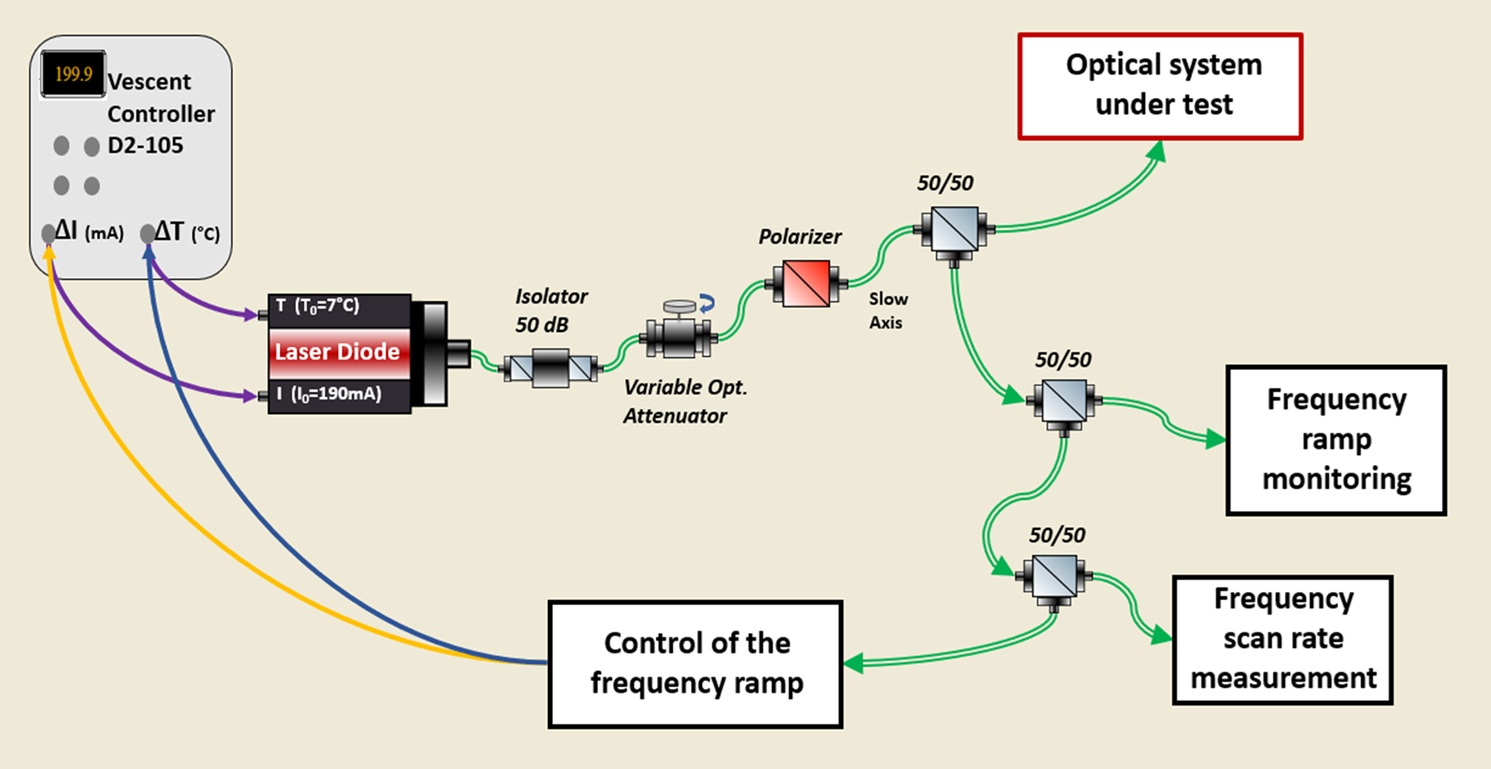}}
\caption[]{Schema of the SL-OGSE prototype}
\label{fig:scheme}
\end{figure}

Figure~\ref{fig:scheme} shows the layout of the SL-OGSE prototype. The laser is a Photodigm laser diode, a current and temperature controller (Vescent D2-105) controls the speed of the laser's optical frequency ramp. An isolator, a variable optical attenuator, and a polarizer are added to the setup before insertion in the DUT.

The system includes a servo unit for the control of the frequency ramp, a unit for the absolute measurement of the frequency scan rate, and a unit for the monitoring of the linearity and reproducibility of the optical frequency.

\section{MEASUREMENTS ON A COMPLEX OPTICAL SYSTEM}

\begin{figure}
    \centering
    \begin{minipage}{0.5\linewidth}
    \includegraphics[width=0.9\linewidth]{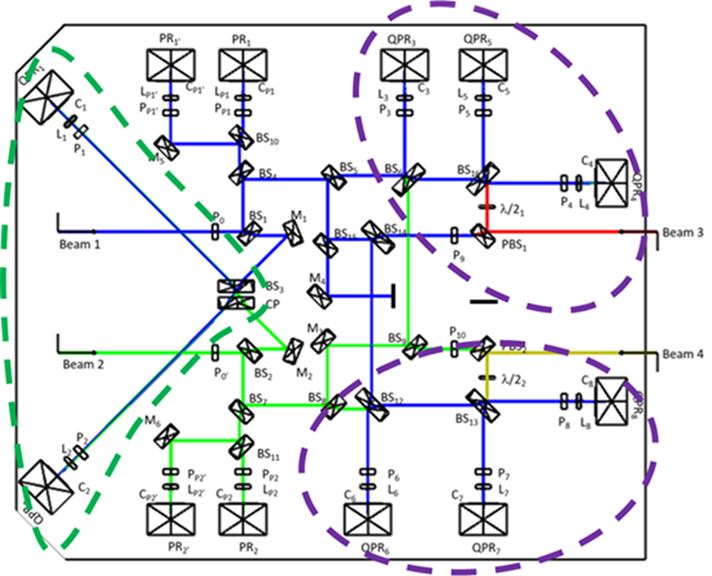}
    \end{minipage}%
    \hfill
    \begin{minipage}{0.5\linewidth}
    \includegraphics[width=0.9\linewidth]{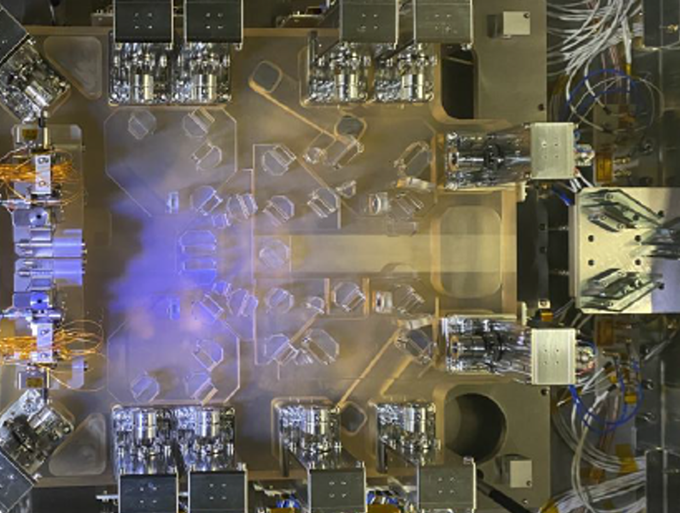}
    \end{minipage}
    \caption{Schema of the ZIFO (left). Dotted purple and green areas indicate the three interferometers. The two laser paths are indicated in blue and green. On the right is a photo of the instrument.}
    \label{fig:zifo}
\end{figure}

We tested the performance of the instrument on a complex interferometer, which was realized to mimic the performance of LISA, the ZIFO (Zerodur Interferometer)~\cite{zifo}. It has been realized at the APC laboratory in Paris and is composed of three interferometers, 8 QPRs (Quadrant Photoreceivers) and 4 SEPRs (Single Element Photo-receivers); see Fig.~\ref{fig:zifo}. We injected the SL-OGSE laser source on two different occasions; to test repeatability of the measurements and also to test some improvements made on the setup along the way.

\section{Data interpretation}

\begin{figure}
\centerline{\includegraphics[width=0.65\linewidth]{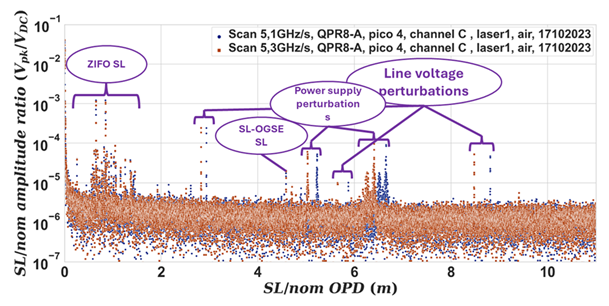}}
\caption[]{Spectrum obtained from a SL-OGSE measurement on the ZIFO bench. There are contributions due to perturbations and defects (electrical grid, SL inside the SL-OGSE) and optical contribution due to the ZIFO design. Measurement is repeated at two different scan rates, 5.1 GHz/s (blue dots) and 5.3 GHz/s (red dots). As a result, a peak for which blue and red dots gather at the same OPD indicates a genuine SL contribution, while contributions that appear as two, blue and red, peaks with a $ \sim 4\% $  difference in OPD are readily identified as foreign perturbations (coupling from the line voltage or from the power supply).}
\label{fig:spectrum}
\end{figure}

\begin{figure}
\centerline{\includegraphics[width=0.65\linewidth]{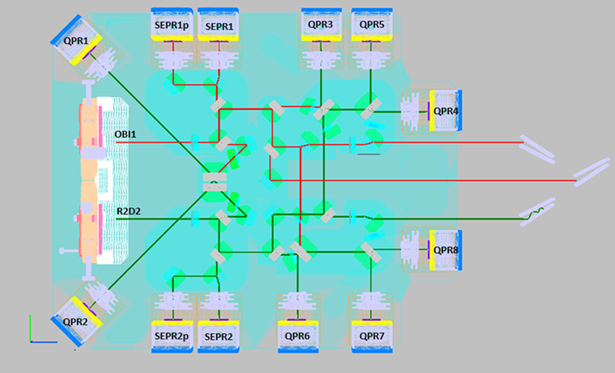}}
\caption[]{ZIFO optomechanical setup used for FRED simulations. The two nominal beams are shown in red and green.}
\label{fig:fred}
\end{figure}

The data obtained during these measurements, see Fig.~\ref{fig:spectrum}, are converted into coordinates Amplitude/OPD associated to each SL peak, and compared with simulations to identify the geometry of the paths on the bench. Simulations are made with ray tracing software FRED (Fig.~\ref{fig:fred}) and provide us with several characteristics of each SL path: optical power, OPL, a picture of the path. We can use this information for comparison with the simulated data.

\begin{figure}
\centerline{\includegraphics[width=1\linewidth]{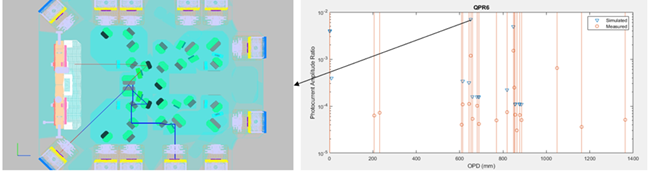}}
\caption[]{Example of comparison between measure and simulation. Right: SL peaks measured with the SL-OGSE are shown as orange dots. Simulated SL paths are represented by blue triangles. For the biggest SL contribution, the corresponding path is shown on the left.}
\label{fig:example}
\end{figure}

As can be seen in the example of Fig.~\ref{fig:example}, there is correspondence in the position of several of the measured points with the simulated points. This allows us to understand the actual path of the rays in the setup and apply corrective measures (in the case of the path of Fig.~\ref{fig:example} the adoption of a beam dump to block it).
This method is powerful but has some limitations:
\begin{itemize}
  \item There is a degree of uncertainty in the estimation of the parameters for the simulation, particularly if the actual optical properties are not known: the simulation will have to be run with the specified AR coating specification, for instance, while the achieved AR coatings are more efficient. If for the optical elements it can be just a moderate difference between the nominal and actual reflection/transmission values, for mechanical or other elements not optically polished, the uncertainty on the R and T coefficient and on the direction of the reflection can be large and lead to false results. This makes it difficult to explain all the measured peaks.
  \item The instrument can measure SL paths with a noise floor which depends on the noise floor in the DUT's photo-receivers, but which can be as low as 1e-6 or better in fractional amplitude (1e-12 in fractional power). At these power levels SL paths are often not clear, less reproducible. Simulations become less reliable. Only above 2e-5 in amplitude accordance simulation/measurement is satisfying.
  \item Because of the uncertainty of some parameters, even when there is accordance between simulations and measurements, experimental verification is advisable to confirm the results.
  \item It is limited by the device under test, if in the setup some deviations from the nominal setup are present, due to mistakes, tolerances etc, simulations will not be able to identify the corresponding SL paths. Also, detector sensitivity and other noises in the DUT can limit performance. Therefore, to really benefit from the accuracy of the instrument, a detailed and accurate knowledge of the DUT is required, otherwise, numerous simulations will be necessary while blindly changing some of the parameters to have some hope to retrieve the SL paths.  This is cumbersome and not always feasible. 
  \end{itemize}
  
Despite these drawbacks we managed to understand the majority of the SL problems the instrument was facing, and the most important contributions have been discovered.

\section{Conclusions}

The SL-OGSE instrument can effectively resolve and measure the contributions to stray light inside a complex optical system and provide the corresponding OPDs. Ray tracing simulations provide a list of modeled stray light paths with fractional power. Matching the two sets of data allows us to identify the stray light paths and offending components, with a view to applying mitigation measures. Deviations from the nominal set-up can result from workmanship errors in the device under test: Then, unless mask/unmask comparisons can be performed significant additional effort may be necessary to comprehensively model the observed contributions.

\section*{Acknowledgments}
This work has been carried on thanks to funding from the CNES and the Region PACA, within the Collaboration LISA France.

\section*{References}
\bibliography{moriond}

\end{document}